\documentclass[sigconf, table, xcdraw]{acmart}

\AtBeginDocument{%
  \providecommand\BibTeX{{%
    \normalfont B\kern-0.5em{\scshape i\kern-0.25em b}\kern-0.8em\TeX}}}

\copyrightyear{2022}
\acmYear{2022}
\setcopyright{acmlicensed}\acmConference[CHI '22]{CHI Conference on Human Factors in Computing Systems}{April 29-May 5, 2022}{New Orleans, LA, USA}
\acmBooktitle{CHI Conference on Human Factors in Computing Systems (CHI '22), April 29-May 5, 2022, New Orleans, LA, USA}
\acmPrice{15.00}
\acmDOI{10.1145/3491102.3501891}
\acmISBN{978-1-4503-9157-3/22/04}

\title[Recommendations for Visualization Recommendations: Exploring Preferences and Priorities in Public Health]{Recommendations for Visualization Recommendations: Exploring Preferences and Priorities in Public Health}

\author{Calvin Bao}
\email{csbao@umd.edu}
\affiliation{\institution{University of Maryland}\city{College Park}\state{MD}\country{United States}}

\author{Siyao Li}
\email{siyaoli@terpmail.umd.edu}
\affiliation{\institution{University of Maryland}\city{College Park}\state{MD}\country{USA}}

\author{Sarah Flores}
\email{sflores3@terpmail.umd.edu}
\affiliation{\institution{University of Maryland}\city{College Park}\state{MD}\country{USA}}

\author{Michael Correll}
\email{mcorrell@tableau.com}
\affiliation{\institution{Tableau Research}\city{Seattle}\state{WA}\country{USA}}

\author{Leilani Battle}
\email{leibatt@cs.washington.edu}
\affiliation{\institution{University of Washington}\city{Seattle}\state{WA}\country{USA}}

\usepackage{subcaption}
\usepackage{caption}

\usepackage{enumitem}
\usepackage{xcolor}
\usepackage{dirtytalk}
\usepackage{multirow}

\newif\ifnotes
\notestrue

\definecolor{highlight}{RGB}{255,0,0}

\newcommand{\revised}[1]{{\color{black}{#1}}}

\ccsdesc[500]{Human-centered computing~Visualization design and evaluation methods}
\ccsdesc[500]{Human-centered computing~Visualization systems and tools}
\ccsdesc[500]{Human-centered computing~User studies}

\keywords{Visualization recommendation systems, algorithmic trust, automation, recommendation source}

\begin{document}

\begin{abstract}

The promise of visualization recommendation systems is that analysts will be automatically provided with relevant and high-quality visualizations that will reduce the work of manual exploration or chart creation. However, little research to date has focused on what analysts \textit{value} in \revised{the design of} visualization recommendations. We interviewed 18 analysts in the public health sector and explored how they made sense of a popular in-domain dataset\footnote{National Health and Nutrition Examination Study 2013-2014~\cite{centers2013nhanes}.} in service of generating visualizations to recommend to others. We also explored how they interacted with a corpus of both automatically- and manually-generated visualization recommendations, with the goal of uncovering how the design values of these analysts are reflected in current visualization recommendation systems. We find that analysts \revised{champion} simple charts with clear takeaways that are nonetheless connected with existing semantic information or domain hypotheses. We conclude by recommending that visualization recommendation designers explore ways of integrating context and expectation into their systems.
\end{abstract}



\maketitle
\section{Introduction}

Data analytics, and especially the creation of informative and useful visualizations of large datasets, can be a time-consuming and complex process. As part of a larger design goal of ``augmenting'' analytics to offload effort to algorithmic systems~\cite{heer2019agency}, there is a growing number of systems that automatically generate and recommend visualizations~\cite{zeng2021evaluation,Zhu2020survey}. Systems such as Voder~\cite{Srinivasan2019augmenting}, Dziban~\cite{Lin2020dziban}, and Tableau Show Me~\cite{mackinlay2007show} can generate visualizations to both surface potentially insightful features of a dataset as well as provide guidance for ``novice investigators'' to generate their own visualizations~\cite{Kwon2011visual}.

The designers of recommendation systems have explicit or implicit \revised{\textbf{design values}} about what charts they surface: for instance, recommenders that purport to automatically surface ``insights''~\cite{Chang2020defining} might place value on particular statistical patterns like outlying values or correlated fields~\cite{demiralp2017foresight,mcnutt2020divining}. However, despite the proliferation of visualization systems in the literature~\cite{zeng2021evaluation,Zhu2020survey}, there has been little work on interrogating these \revised{design values}, and observing matches and mismatches between the values of recommendation system \textit{designers} and \textit{consumers}. Our work is therefore focused on a central question: \textbf{what sort of visualizations do people \textit{want} to see, and how well do these preferences actually \textit{align} with the sorts of visualizations that algorithmic recommendation systems currently provide?} 

Prior work considers how people react to different recommendation sources~\cite{Peck2019Data,Zehrung2021vis}, but does not consider the \emph{priorities} and \emph{expectations} analysts have when creating their own recommendations for other analysts.
Without a deeper understanding of how analysts themselves think about the visualization recommendation process, new recommendation engines may barely help~\cite{zeng2021evaluation}, and possibly even hinder~\cite{correll2019ethical}, an analyst's ability to explore their data, creating a ``double-edged sword''~\cite{li2021exploring} of potentially ``opaque, inflexible, brittle, and domineering''~\cite{mcnutt2020divining} analysis.

In this paper, we present the results of a pre-registered\footnote{\url{	https://aspredicted.org/AEI_GBA}} qualitative study designed to interrogate and elicit \revised{design values} around generating and evaluating visualization recommendations. Our study, conducted with public health researchers supplied with a sample dataset of U.S. self-reported health data, consists of two components:
\begin{enumerate}
    \item an \textbf{ideation task} where participants, with the help of an experienced visualization designer working with Tableau, sketched and then realized their own visualization recommendations for an imagined client seeking to influence public policy, and
    \item a \textbf{selection and ranking task} where participants explored a gallery of recommendations (some generated automatically by systems, and some by human curators) and selected the ones they felt were most valuable for their client.
\end{enumerate}

We chose these tasks and this participant pool to examine points of friction between the values of recommendation systems and analysts. \revised{I.e.,} we wanted to contrast the (often) domain-agnostic assumptions of visualization recommendation systems with the specific domain expertise and context of our participants, and contrast the (frequent) focus on narrowly defined statistical findings in recommendation systems with the unconstrained and diverse rhetorical and persuasive goals of our participants.

Of the \revised{design values} we encountered in our exploration, the three most prominent that our participants valued in recommended visualizations were:
\begin{enumerate}
    \item \textbf{simplicity}--- participants, often with an assumed audience in mind, valued simple visualization designs over more complex ones, and visualizations with one clear takeaway over more nuanced or complex data stories. Titles and labels, filtering, and aggregation were common strategies to reduce the complexity of data.
    \item \textbf{relevance}--- in addition to a preference for the removal of extraneous data from recommendations, participants also made efforts to tailor their charts to their domain of interest. E.g., a preference for bivariate visualizations with anticipated casual relationships (e.g. that one variable would ``drive'' another, or produce a clear ``trend'').
    \item \textbf{interestingness}--- participants were reluctant to provide visualizations that failed to show clear trends, \revised{group} differences, or other strong signals. Participants wanted recommended charts to provide direct evidence for or against particular hypotheses, or to promote specific follow-up actions.
\end{enumerate}

These \revised{design values} suggest both \textit{opportunities} and \textit{dangers} for designers of future visualization systems. On the one hand, they suggest benefits for incorporating additional data semantics or explicit user intent into recommendation systems to better meet the goals of analysts. On the other hand, they suggest that care should be taken to communicate complex or ambiguous trends in the data that might arise in the recommendation process, and that the desire to surface strong signals promotes a form of exploratory data analysis that lends itself to false positives or other dangers~\cite{pu2018garden,zgraggen2018investigating} to the reliability or robustness of findings.

\label{sec:introduction}

\section{Related Work}
\label{sec:related-work}

Our research questions and experimental design are informed by  assumptions and goals behind the design of existing visualization recommendation systems, as well as by prior studies that involve participants expressing their preferences amongst visualizations from heterogeneous sources or creating novel and heterogeneous visualizations themselves. We therefore highlight three topics of related research: visualization recommendation systems, assessment of those systems (and visualizations in general), and visualization construction for novice users.

\subsection{Visualization Recommendations}
\label{sec:related-work:recommendation}

Visualization recommendation systems aim to ease the process of visualization authoring or exploratory data analysis for different user groups~\cite{Zhu2020survey,Kaur2017review}. Each system has its own set of \revised{metrics} and structures to represent what users find valuable to visualize in a dataset~\cite{zeng2021evaluation}. \revised{For example, some recommendation systems prioritize perceptually effective encoding channels for a given set of data attributes (e.g., \cite{Lin2020dziban,mackinlay2007show,mackinlay1986automating}), popular visualization designs that other users have created in the past (e.g., \cite{Hu2019Vizml}), or specific types of data trends such as pairwise correlations between attributes~\cite{demiralp2017foresight} or significant differences among sub-populations in the dataset~\cite{vartak2015seedb}. However, these priorities are often set in a way that is agnostic to either the domain of interest or the particular analytical goals of the user. When the system's and user's priorities are misaligned, the system may generate distracting and ineffective recommendations~\cite{zeng2021evaluation}. In this paper, we seek to clarify what analysts prioritize when designing their own recommendations in the context of public health, and to understand how analysts' priorities compare with those of existing systems.}

We divide \revised{existing visualization recommendation} systems into three categories: Auto-Insight, Encoding, and Q\&A, although we note that these categories are not necessarily mutually exclusive, and that recommendation systems can and do incorporate \revised{design values} or patterns from multiple modalities.

\subsubsection{Auto-Insight}
\label{sec:related-work:autoinsight}
Auto-insight systems automatically detect and visualize meaningful attributes, trends, or other statistical properties within a provided dataset~\cite{li2021exploring},
removing some of the labor or luck involved in manual exploratory data analysis~\cite{heer2019agency}.
These data insights can be given in the form of text describing statistical patterns in the data or through visualizations~\cite{Law2020Characterizing}. 
Example systems include Voder \cite{Srinivasan2019augmenting}, which focuses on textual facts and insights, as well as PowerBI Quick Insights~\cite{Mihart2021Types}, Foresight~\cite{demiralp2017foresight}, and Amazon QuickSight~\cite{AmazonQuickSight}, which focus on \revised{insights presented as visualizations}. Voder creates textual ``data facts'' based on the dataset's attributes to assist users in interpreting generated data visualizations and communicating findings~\cite{Srinivasan2019augmenting}. PowerBI's Quick Insight~\cite{Mihart2021Types} panel searches through different subsets of a dataset and detects particular classes of statistical features (e.g., outliers, variance, correlation, categories with a strong majority) to generate insights during data exploration. Similarly, Foresight~\cite{demiralp2017foresight} ranks visualizations based on statistical properties present in the data. Lastly, Amazon QuickSight~\cite{AmazonQuickSight}, within the broader Amazon Web Services ecosystem, also creates data summaries using in-house algorithms, allowing users to upload and integrate their own models.


Although each of these systems provides a different means of exploring data and communicating insights, they generally lack explicit explanations for how insights are generated, leading some users to distrust the results~\cite{zeng2021evaluation}. Furthermore, users do not necessarily know whether these recommendations cover everything that could or should be learned from the given dataset~\cite{Law2020Characterizing}. This lack of transparency and consideration of user context (i.e., user preferences and intended recommendation goals) may result in bias, unreliability, and disruption of the exploratory data analysis process~\cite{li2021exploring,Zehrung2021vis}. Another worry is that, by exhaustively searching for potentially interesting patterns, auto-insight systems can function as ``p-hacking machines''~\cite{correll2019ethical,pu2018garden}, surfacing ``insights'' that are ultimately spurious or misleading.

Our experimental design is most closely aligned with the goals and values of auto-insight recommenders, in that our participants were asked to generate meaningful visualizations for their clients without constraints on fields or designs of interest, although we note overlaps with other forms of recommenders below.

\subsubsection{Encoding}
We define encoding recommendation systems as those that suggest designs of individual visualizations given user-specified data attributes. These recommendation(s) are often based on the characteristics of the data and expert knowledge on the expressiveness and effectiveness of different encoding channels or chart designs~\cite{mackinlay1986automating}. 


These systems employ a variety of approaches in how they encode expert knowledge. Draco~\cite{Moritz2019Formalizing} uses a set of constraints to assist users in visualization design and prioritize visual exploration, promoting effective encodings, and predicting the best visualization through a ranking system. Dziban~\cite{Lin2020dziban} builds upon the Draco knowledge base while incorporating chart similarity logic to create a balance between ``automated suggestions and user intent.'' Show Me~\cite{mackinlay2007show} either suggests (and automatically generates) particular chart designs given the data types of selected data attributes, or allows the user to progressively construct a chart by adding attributes one at a time, automatically suggesting new encodings or chart designs. Graphscape~\cite{Kim2017Graphscape} uses a directed graph model in which nodes represent chart specifications and edges represent transitions between charts. This model enables Graphscape to recommend alternative designs by minimizing the perceptual distance between new recommendations and visualizations previously seen. Table2Charts~\cite{Zhou2021Table2charts} takes table-chart pairs and learns patterns that assist in generating recommendations. Several systems also focus on multi-dataset exploration. For example, GEViTRec~\cite{Crisan2020Gevitrec} recommends visualizations across multiple datasets by looking for linking fields or domain-centric constraints. Lastly, Data2Vis~\cite{dibia2019data2vis} uses neural networks to translate a given dataset into a resulting visualization specification, based on a training set of presumably well-designed Vega-Lite~\cite{satyanarayan2017vega-lite} specifications. VizML~\cite{Hu2019Vizml} applies a similar learning approach to Plotly visualizations.

While our experimental framing was less aligned with the \revised{design values} of these systems, as our participants had free choice over which variables to include, the ability of our participants to create and select their own designs, and to iteratively alter the default designs generated over the course of the experiment, allowed us to see if existing assumptions around expressiveness and effectiveness matched the preferences \revised{and priorities} of our participants (who, while embedded in their domain of interest, had varying levels of expertise in visualization design).

\subsubsection{Q\&A systems}
\label{sec:related-work:qa}
We consider Q\&A recommenders to be systems where the recommendation engine and the user can engage in one or more rounds of communication for the generation and refinement of recommendations. A prototypical Q\&A system might take as input textual questions, suggestions, and/or attributes (``questions'') and produce as output an appropriate visualization (an ``answer''). Amazon QuickSight and Tableau Ask Data are examples of systems that provide this feature. QuickSight~\cite{AmazonQuickSight}, for instance, features a search bar wherein users can enter natural language questions about their data. The user's intent is then inferred from the questions, and the system returns an answer in the form of a number, visualization, or table. Tableau's Ask Data~\cite{Wesley2011analytic}, a system for performing ad-hoc exploration and analysis, also incorporates natural language interaction features: users type in natural statements or questions into an input bar and the system produces a chart~\cite{tory2019do}.

Although the user can explicitly tell these Q\&A systems what they are interested in, the extent to which systems are truly cognizant of or reactive to the intent of the user is often unclear~\cite{tory2019do}. For example, a user may have a chart in mind when asking a question (or selecting attributes) but receive an entirely different chart as output, deviating from their expectations. How should systems respond to ambiguous questions from the user?~\cite{hearst2019toward}

In our study, we are interested in understanding what users generally value in \revised{the design and construction of visualizations}, which can inform general purpose guidelines for creating intent-focused Q\&A systems, and other visualization recommendation systems. Our study protocol also allowed participants to iterate with us to refine their recommendations, affording an analysis of what sorts of refinement or repair operations are common in visualization recommendation with human partners, that could similarly be of use for designers of automated Q\&A systems.

\subsection{Visualization Recommendation Assessment}
\label{sec:related-work:assessment}

While a full consideration of all of the ways visualizations can and have been assessed is out of the scope of this work (see Lam et al.~\cite{lam2011empirical} for a typology), we focus on studies dealing with eliciting preferences from sets of unfamiliar visualizations that have been presented to participants, especially in the context of \textit{recommendation}. Peck et al.~\cite{Peck2019Data} performed a qualitative study where participants were asked to assess their attitudes towards an array of infographics, highlighting how different beliefs and stances can influence the perceived quality, utility, and trustworthiness of a visualization. Lee et al.~\cite{Lee2016How} explored the process of making sense of unfamiliar visualizations through a think-aloud procedure similar to the one we adopt, with an emphasis on investigating what factors influence the interpretability of a visualization. In our study, we look to these works as a model for exploring attitudes towards \textit{existing} visualizations, but we also include a visualization \textit{authoring} step in order to assess specific design characteristics our participants valued when creating their own recommendations.

Another paper we use as a model is Zehrung and Singhal et al.~\cite{Zehrung2021vis}, where participants were given sets of recommendations from an unseen visualization recommender, with the specific goal of evaluating how the stated provenance of the recommender (human or algorithmic) impacted perceived quality and trust. While we similarly intermix human and algorithmic recommendations in our study with the goal of investigating any systematic differences between the two sources, our work moves away from questions of trust and provenance and towards broader issues of perceived utility and impact. Lastly, Zeng et al.~\cite{zeng2021evaluation} propose a framework for specifying multiple visualization recommendation algorithms within the same semantic space to enable quantitative comparison and evaluation. While our mixture of qualitative and quantitative methods complicate this process, our study findings could be incorporated into the Zeng et al. framework to improve evaluation of end-user preferences and expectations for recommended visualizations.

\subsection{Barriers and Methods for Eliciting Visualizations }
\label{sec:related-work:construction}

As we intended for our participants  to both evaluate existing visualization recommendations \emph{and} generate their own, we explored potential processes and pitfalls for eliciting visualizations from diverse audiences, especially audiences who may lack experience with existing visualization design tools. Several visualization construction barriers exist for visualization users (especially novices) --- Grammel et al.~\cite{Grammel2010how} finds that novices struggle with navigating and mapping the relationships between visualization concepts: exploratory questions, data attributes, and visualizations during the construction process. Other barriers include several reported by Kwon et al.~\cite{Kwon2011visual}: a failure to interpret visualizations properly and a failure to match expectations and functionality of the visualization. These barriers often caused frustrations among novice visualizers. This presents an underlying ``gulf of execution'' between the types of visualizations that users want versus what visualization recommendation systems actually generate. For this work, we are particularly interested in how more advanced analytics users approach this gulf.

To help reduce the barriers to effective visualization construction, free-form sketching can serve as an expressive medium of converting internal thought to external representations ~\cite{Walny2015exploratory,Kirsh2010Thinking}. Moreover, work by Tversky highlights the power of the sketching process to reveal the designer's underlying ideas and reflect core aspects of one's prioritization~\cite{Tversky2008Making}, a power that is used by systems such as SketchStory~\cite{Lee2013sketchstory} for fluid and flexible visualization authoring. A tangential effect of free-form sketching is that it provides direct interaction. Studies on whiteboard usage showed how whiteboard sketching enables people to immediately externalize ideas without being interrupted by or having to translate their ideas to another medium or system~\cite{Walny2011Visual}. We incorporated free-form sketching into our study to reduce the construction complexity for our participants and to better observe expectations for the visualizations they create.

\subsection{Summary}
\label{sec:related-work:summary}

Our analysis of prior work points to a wide space of visualization recommendation systems that nevertheless prioritize specific statistical features, low-level analysis tasks, and visualization design rules, all of which have advantages in particular scenarios, but may or may not capture the specific \revised{priorities} and mental models of analysts more broadly. Prior work also suggests relevant strategies for working with audiences across levels of data expertise or engagement to develop rich frameworks around understanding, values, and priorities in visualization. Our study seeks to integrate these two perspectives by performing a human-centric assessment of the priorities and values of visualization recommendations, the results of which can guide the designers of future visualization recommendation systems.



\section{Motivation}
\label{sec:motivation}

Our study is motivated by a potential gap in \revised{\textit{design values}}: between the values of designers of visualization recommendation systems (who might prioritize highlighting a particular subset of statistical patterns, data facts, or ``insights'') and those of human analysts (who might have more semantically rich or teleological expectations of their visualizations).
With a deeper understanding of what analysts prioritize as they create and rank visualizations for later recommendation, we can compare our observations with how visualization recommendation systems are currently designed, and provide concrete feedback for how current and future systems can be refined to more closely align with the goals and values of their end-users.

We break our broader research question (\emph{what do analysts value in \revised{the design of} visualization recommendations, and are these \revised{design values} reflected in current visualization recommendation systems?}) down into three sub-questions to investigate through our study:
\begin{itemize}[nosep]
\item \textbf{RQ1}: What characteristics of a visualization design do analysts prioritize when recommending them to colleagues?
\item \textbf{RQ2}: What do analysts prioritize when evaluating visualization recommendations from other sources?
\item \textbf{RQ3}: How do the recommendations made by analysts align with those created from other sources in terms of visual form or analytical purpose?
\end{itemize}

While recommendation systems are often agnostic or insensitive to data domain or analytic intent, our belief is that the perceived usefulness of a visualization is often task- and domain-dependent~\cite{sedlmair2012design}. For these reasons, we focus on a single domain in this work in order to specifically elicit any potential tensions between the domain insensitivity of many automatic recommenders and the domain knowledge and intents of our participant pool. Specifically, we investigate how researchers and professional analysts working in the public health sector create and evaluate visualization recommendations for a goal of presenting information to shape public policy. 

Our questions are ones of exploring or enumerating \textit{alignment}  \revised{in design values} rather than evaluating predictions or building models. As such, we do not enumerate hypotheses for testing, but focus more on descriptive quantitative reports of our findings augmented with qualitative data.


\section{Experiment Design}
\label{sec:experiment-design}

We designed a pre-registered\footnote{\url{https://aspredicted.org/7d7gd.pdf}} experiment to better understand what visual or data characteristics analysts prioritize when creating visualizations for other analysts, and what analysts purport to value when presented with a gallery of human-curated and algorithmically-generated recommendations.


To accommodate a wider range of participants as well as to abide by COVID-19 pandemic protocols, the study was conducted online using the video conferencing platform Zoom. Participants shared their screen with the experimenters, and completed the study using Google Jamboard, an online sketching and whiteboard tool. In the following subsections, we describe our participant pool, pilot study, and visualization artifacts used for the experiment, and then walk through the entirety of an interview, describing each phase and how the participant was to interact with the interviewing team.

Additional study details, including transcripts, sketches, generated analyzes, data tables, and analyses are available at
\revised{\url{https://osf.io/xeub3/}.}

\subsection{Participants}
\label{sec:experiment-design:participants}
After approval by our institutional IRB, we recruited 18 participants through a combination of university mailing lists, snowball sampling through research collaborators, and advertising on social media. We employed different methods of sampling to broaden population groups of participants to minimize the selection bias in our recruitment process.

\revised{We present demographic information about our participants in \autoref{tab:demographic-info}.} Our participants ranged between 18-64 years old, with \revised{two being between 19-24, ten being between 25-34 years old, three being 35-44 years old, and three being 45+ years old}. \revised{In terms of domain expertise, at the time of study, four participants were current graduate students and the remaining fourteen participants were working as public health professionals in various capacities, ranging in roles from project director, health program administrator, faculty member, and research scientist. Regarding frequency of creating data visualizations, nine }participants reported creating visualizations at least once a month, five reported creating visualizations at least once a week, and three reported creating visualizations daily. The remaining participant reported creating visualizations rarely (less than once a month). To qualify for participation, participants had to have at least two years of \revised{industry analyst or research experience} in public health. We compensated participants with a \$25 Amazon gift card for completing the study.

By the end of the study, we collected a set of 53 visualization sketches (all participants but one sketched out three, while the one sketched only two) and 18 rankings of the visualizations in a gallery of recommendations from our participants.


\begin{table*}[]
\caption{\revised{Demographic information about each study participant, labeled by participant ID (PID).}}
\revised{
\begin{tabular}{llllll}
\hline
\textbf{PID} & \textbf{Age Group} & \textbf{Gender} & \textbf{Job Title}                                                     & \textbf{Perform Data Analysis} & \textbf{Create Visualizations} \\ \hline
\rowcolor[HTML]{EFEFEF} 
1            & 19 - 24            & Female          & Consultant                                                             & Weekly                         & At least once/month            \\
2            & 25 - 34            & Female          & Research Coordinator                                                   & Weekly                         & At least once/week             \\
\rowcolor[HTML]{EFEFEF} 
3            & 25 - 34            & Female          & Graduate Student                                                       & \textless Once/month           & At least once/month            \\
4            & 25 - 34            & Female          & Project Coordinator                                                    & \textless Once/month           & At least once/month            \\
\rowcolor[HTML]{EFEFEF} 
5            & 35 - 44            & Female          & Consultant                                                             & Weekly                         & At least once/week             \\
6            & 25 - 34            & Male            & Research Coordinator                                                   & \textless Once/week            & At least once/week             \\
\rowcolor[HTML]{EFEFEF} 
8            & 19 - 24            & Female          & \begin{tabular}[c]{@{}l@{}}Health Program\\ Administrator\end{tabular} & \textless Once/month           & At least once/month            \\
9            & 45 - 54            & Female          & Assistant Professor                                                    & Weekly                         & At least once/month            \\
\rowcolor[HTML]{EFEFEF} 
11           & 25 - 34            & Female          & Graduate Student                                                       & \textless Once/week            & At least once/month            \\
12           & 25 - 34            & Female          & Research Scientist                                                     & Daily                          & Daily                          \\
\rowcolor[HTML]{EFEFEF} 
13           & 55 - 64            & Female          & Lecturer                                                               & \textless Once/week            & At least once/month            \\
14           & 35 - 44            & Female          & Faculty                                                                & Weekly                         & At least once/month            \\
\rowcolor[HTML]{EFEFEF} 
15           & 45 - 54            & Female          & Project Director                                                       & \textless Once/week            & At least once/month            \\
16           & 25 - 34            & Female          & Research Scientist                                                     & Daily                          & Daily                          \\
\rowcolor[HTML]{EFEFEF} 
17           & 35 - 44            & Male            & Project Coordinator                                                    & \textless Once/week            & At least once/week             \\
18           & 25 - 34            & Female          & Faculty Specialist                                                     & Weekly                         & At least once/week             \\
\rowcolor[HTML]{EFEFEF} 
19           & 25 - 34            & Female          & Graduate Student                                                       & \textless Once/month           & At least once/month            \\
20           & 25 - 34            & Male            & Database Administrator                                                 & \textless Once/month           & Daily                          \\ \hline
\end{tabular}
}
\label{tab:demographic-info}
\end{table*}

\subsection{Experimental Dataset}
\label{sec:experiment-design:dataset}

To ensure that we selected and presented data that aligned with the interests of our target participants, we solicited feedback from experts in public health at our primary authors' home institution. These experts provided guidance on relevant datasets, attributes that would be of particular interest to a public health audience, and which groups and departments to target for recruitment. Based on this feedback, we selected a vertical subset of the National Health and Nutrition Examination Survey (\emph{NHANES}) from 2013-2014 for use in our study~\cite{centers2013nhanes}. Twenty attributes were extracted from the \emph{NHANES} dataset, covering the \emph{Demographics, Examinations, Dietary,} and \emph{Questionnaire} response categories. We randomized the ordering of attributes for each participant to mitigate order effects. A sample table of 6 records was provided to participants, so they could see the available attributes and their data types. Participants were also given the option to view the dataset in its entirety through an online link to a spreadsheet.





\subsection{Pilot Study}
\label{sec:experiment-design:pilot-study}

We conducted an initial pilot experiment with five participants. We also presented our experimental protocol to two faculty members at our institution's School of Public Health for additional feedback. 

Initially, we asked pilot participants to hypothesize about which attributes would be important for visualization. This helped us to narrow down our list of data attributes that we believed to be valuable and relevant to in-domain analysts. We specifically extracted only these data attributes from the broader dataset. We presented this subset to each pilot participant and asked them to sketch five visualizations they would recommend to other analysts that would explore the same dataset. We found that asking for five sketches prohibitively extended the length of the study, as they reported that it was difficult to create five sufficiently distinct and interesting visualizations in the allotted time. We decreased the number to three, resulting in a final approximate study length of 60 minutes. We validated this updated study design with an additional pilot participant. An additional modification as a result of this piloting was to allow both solicitation of sketches via Jamboard (which some participants found limiting or difficult to use) as well as via hand-drawn sketches emailed directly to the experimenters.

We also used the pilot study to seed the gallery of visualizations that we ultimately used for our selection/ranking task in the final experiment. Our gallery was created via a mixture of the visualizations created by our pilot participants as a result of their ideation task (5 visualizations) and the visualizations favored by our pilot participants from a gallery of visualizations generated by recommendations systems Dziban~\cite{Lin2020dziban}, Voyager~\cite{Wongsuphasawat2017Voyager}, PowerBI~\cite{Mihart2021Types}, Tableau ShowMe~\cite{mackinlay2007show}, and Amazon QuickSight~\cite{AmazonQuickSight}. We selected five of the algorithmically-recommended visualizations with the most votes from our pilot participants to add to our finalized gallery of recommendations: the resulting favorite visualizations came from Dziban (3 visualizations) and Voyager (2 visualizations).



\subsection{Experiment Flow}
\label{sec:experiment-design:experiment-flow}

Our study consisted of four phases:

\begin{itemize}[nosep]
    \itemsep0em
    \item \textbf{Phase 1. Tutorial} The participant receives a brief tutorial on how we use Jamboard to conduct the virtual experiment. The participant was then introduced to the task for Phases 2 and 3, and given an opportunity to review the dataset.
    \item \textbf{Phase 2. Visualization Ideation with Mediator\revised{ and Wizard}} The participant sketches 3 visualizations based on the given dataset and task\revised{ in conversation with a mediator.  A ``wizard'' then translates these sketches into final visualizations using Tableau Desktop.}
    \item \textbf{Phase 3. Visualization Selection and Ranking} The participant selects and ranks 5 visualizations from a set of 10 visualizations (5 human-generated and 5 system-generated).
    \item \textbf{Phase 4. Post-survey} Participants complete a short survey  to provide closing comments and demographic information.
\end{itemize}


\subsubsection{Tutorial}
\label{sec:experiment-design:tutorial}
We explained the features available on Jamboard (e.g., pencil, eraser, drawing shapes) Participants spent five minutes familiarizing themselves with Jamboard's features. They then were given the task, which is as follows: 

\say{\emph{You have been paired with an analyst to develop slides to present to a client. The client is developing public policy to improve health outcomes in the US. Create a set of 3 visualizations (a.k.a. charts and graphs) that you would recommend to this client.}}

They were then given an opportunity to explore and review the dataset, with minimal input from the interviewers.

\subsubsection{Visualization Ideation with Mediator \revised{and Wizard}}
\label{sec:experiment-design:wizard-of-oz}
Participants were then asked to select and rank a short list of attributes of interest to them, to better ground them within the task. They then were asked to sketch visualizations while following a think-aloud manner. The participant shared with the mediator their sketches and any further necessary visualization specifications, while discussing the ideas, goals, and motivations behind the visualization. Then, the \revised{wizard} re-created the participants' sketch with visualization software, Tableau Desktop, similarly to the process used by Grammel et al.~\cite{vis_novices} in their study. \revised{During this sub-phase, the mediator guided the participant by explaining set tasks, and organizing the jamboard. Concurrently, the wizard would share their Tableau visualizations with the participant, receiving feedback and then updating the visualizations based on this feedback to ensure that the final designs matches the participant's sketched intentions.}

\subsubsection{Visualization Selection and Ranking}
\label{sec:experiment-design:selection}

\begin{figure*}[htp]
    \centering
    \includegraphics[width=0.95\textwidth]{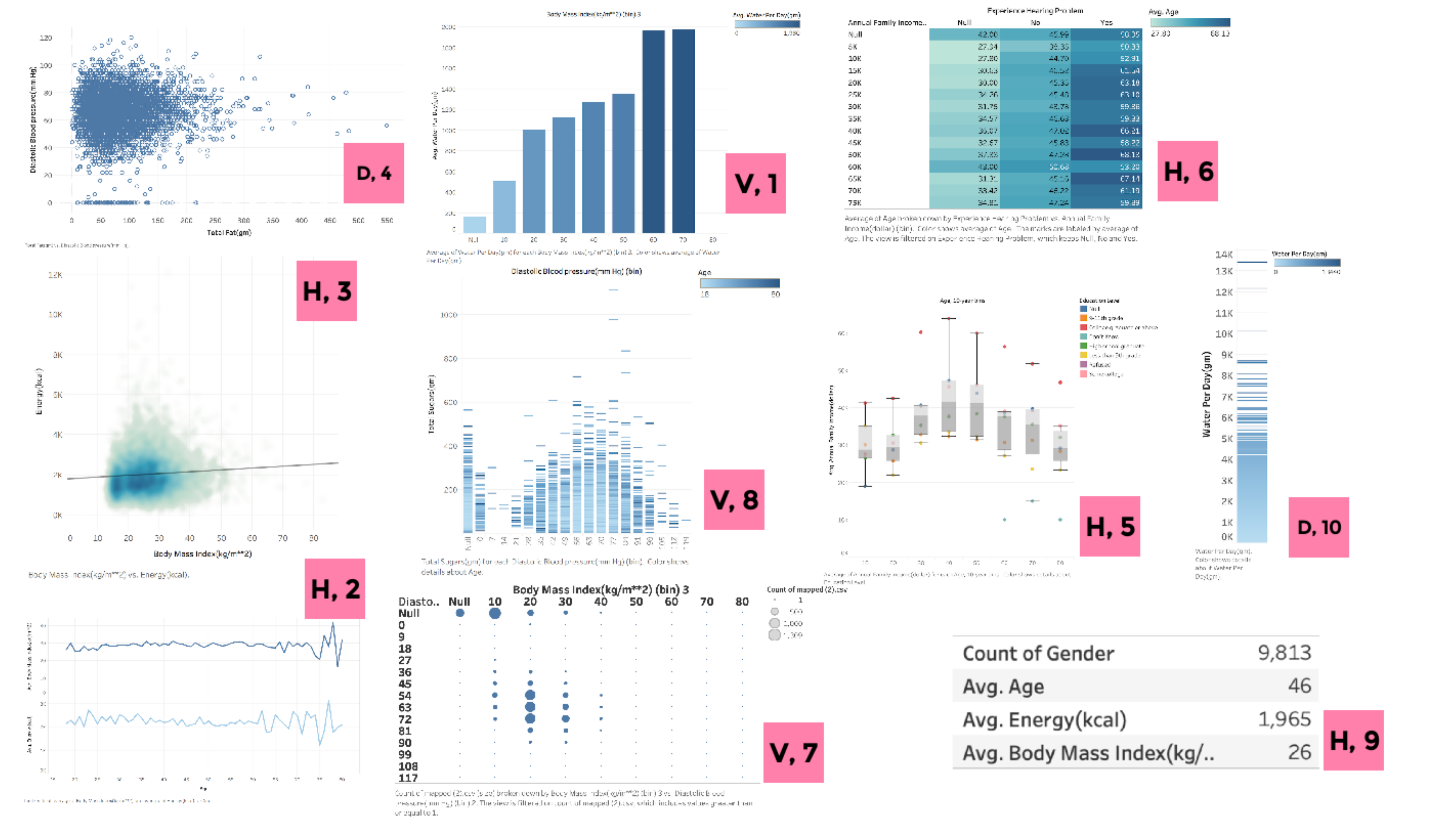}
    \caption{The gallery of visualization recommendations provided to our study participants. While we did not share the origin of these visualizations with our participants, for illustration in this paper we have labelled the chart provenance: those marked ``H'' were generated by pilot participants as part of their ideation sessions, those labelled ``V'' were recommended by Voyager~\protect\cite{Wongsuphasawat2017Voyager}, and those labelled ``D'' were recommended by Dziban~\protect\cite{Lin2020dziban}. We also include the overall rank of the recommendation, according to our participants. Overall rank is determined by summing reverse ranking data grouped by the visualization ID, then sorting in descending order.}
    \label{fig:experiment-design:gallery}
\end{figure*}
Afterward, participants were presented with a gallery of 10 visualizations (\autoref{fig:experiment-design:gallery}, a mixture of human-recommended and algorithmically recommended visualizations (see \autoref{sec:experiment-design:pilot-study})). We did not specify the provenance of these visualizations. The display order of visualizations was randomized for each participant to account for order effects. Participants were asked \revised{by the mediator} to select and rank the top 5 visualizations within this set that they favored and would recommend to others in relation to the prompt given in \ref{sec:experiment-design:wizard-of-oz}.

\subsubsection{Post-survey}
\label{sec:experiment-design:survey}

The last step of the experiment was for participants to fill out a short post-interview survey about their visualization experience, data analysis experience, age, and work experience, e.g., their highest level of education, years of experience, prior statistical experience, current job title, and also the frequency for performing data analysis.

\subsubsection{Data Collection}
\label{sec:experiment-design:data-collection}
Participants' survey responses were collected using online forms and stored as CSV files. Participants' sketches (uploaded images and Jamboard designs), gallery selections and survey responses were also collected. The sketches were translated into corresponding Vega-Lite specifications~\cite{satyanarayan2017vega-lite}. The gallery selections and rankings were also stored as CSV file. Participants shared their screen during the experiment and screen capture was recorded, providing video and audio recording of each interview session. Each session was transcribed. Experimenters also took notes during the experiment. \revised{To preserve participant privacy we do not share these audio or video records directly, but only the transcripts, notes, and codes, and after a manual process of redacting identifying information. We share this }collected data as a public resource on OSF:
\revised{\url{https://osf.io/xeub3/}.}

\subsection{Experiment Design Limitations \& Trade-offs}
\label{sec:experiment-design:limitations}

We considered multiple trade-offs in the design of our study, which we discuss here.

\subsubsection{Limitations in Data Collection}
In our experiment, we chose to reduce the number of attributes presented to participants for two reasons related to participant accommodation.
First, we found in our pilot study that having access to the entire \emph{NHANES} dataset was overwhelming for participants, and they reported encountering ``analysis paralysis'' when deciding what attributes to select and what visualizations to sketch.
Second, with the original number of attributes to choose from, thousands of attribute combinations were possible for creating visualizations, which could have led to participants having little or no overlap in their attribute and visualization preferences.
Though this observation could hint at the need to implement scope-reduction (in terms of attributes) within visualization recommendation systems, for logistical purposes, we decided to present just a subset that was approved by our pilot participants.

We also reduced the number of sketches we asked participants to create in the Visualization Ideation Phase of the study. We made this decision based on study length as observed through our pilot (see \autoref{sec:experiment-design:pilot-study}). 

\subsubsection{Limitations of our Participant Pool}
Though visualization recommendation systems are of broad interest for analysts across many domains, we recruited participants specifically with experience in analyzing public health data, resulting in a relatively specified participant pool. Our findings may not necessarily translate to other domains. However, given the importance of visualization in communicating public health information, and the richness of our selected dataset (\emph{NHANES}), we believe our experiment still provides meaningful insights for the visualization community. We also believe that our decision to focus on participants with particular domain expertise would serve to highlight discrepancies in their design values compared to the relatively domain-agnostic priorities of algorithmic visualization recommendation systems.

\revised{We focused on participants with domain expertise who all expressed at least some familiarity with the sort of survey data that we used in our experiment. We expect that entirely different design challenges and trade-offs would result among participant pools with varying levels of familiarity with the source data.}

\subsubsection{Limitations of a Remote Study}
Due to limitations imposed by the COVID-19 pandemic, we conducted all interviews remotely, with telecommunication software (Zoom) and an online sketching tool (Google Jamboard). Given that participants can vary widely in  their experience with online tools, this mode of administering the experiment likely impacts our findings. That being said, our participants were able to successfully complete the study, and share meaningful visualization recommendations. When studies can be safely conducted in-person again, an in-person study would enable us to administer the study without the limitations imposed by our software.


\subsubsection{Limitations of the Mediation Approach}
As observed in prior studies~\cite{Grammel2010how,tory2019do}, the human mediation (or \say{Wizard-of-Oz}) process can sometimes lead to confusion on the part of participants, given the interplay between a participant's intent, communication of this intent to the ``wizard,'' and what can be feasibly designed by the participant using real data.
Many times, participants were surprised to see the result of the visualization after their initial sketch. We observed that most of these surprises were due to a mismatch in expectations between the user's understanding of the data, and what the data actually supported.

Furthermore, we acknowledge that since the participant's intent is being interpreted by the mediator, our recreations of participants' sketches may not perfectly match their expectations. The creation of visualization recommendations under longer time frames, with deeper analyses of the data, and with additional rounds of feedback we suspect would generate categorically different sorts of visualization recommendations, but this remains an area of future study.

\section{Analysis}
\label{sec:analysis}

To reiterate, the main question driving our research is: \emph{what do analysts value in visualization recommendations, and are these values reflected in the design of current visualization recommendation systems?}
To answer this question, we collected study data regarding participants' design values and preferences as they created their own visualization recommendations and ranked visualizations from a pre-defined gallery of recommendations.
First, we describe how we qualitatively and quantitatively analyzed this data to shed light on the perspectives and design values of analysts from the public health sector when recommending visualizations.
Then, we present our analysis results, organized around the research questions listed in \autoref{sec:motivation}. All of our analysis code and results are shared in our open-source repository on OSF:
\revised{\url{https://osf.io/xeub3/}.}

\subsection{Qualitative Analysis Methods}

\revised{We adoped a grounded theory-inspired approach for qualitatively analyzing our study data.} Given our collected data (transcriptions, experimenter notes, participant sketches, and participant rankings), we used emergent coding to identify common themes in how participants design and rank visualization recommendations. \revised{In this way, we could avoid making assumptions a priori about our participants and thereby maximize our ability to observe a broad array of participant preferences.} As part of our coding process, we explored potential themes regarding the visual forms of the visualizations (e.g., what encodings or visualization types our participants selected) as well as each participant's analytical intent (e.g., the types of statistical patterns our participants wanted to observe) and semantic context (e.g., the domain knowledge applied to value one visualization over another).


As a starting point, four members of the team independently went through differing subsets of at least 6 participants of interview data and created a set of themes and codes based off their assigned participants. Assignments included overlapping subsets, to not only improve saturation but also to validate that synthesized ideas were consistent for the same participant. After all sets of notes, themes, and codes were completed, one member of the team compiled \revised{an ``aggregate''} codebook based off the themes \revised{derived from each set. Similar themes were merged, grouping together notable quotes from participants, while broader themes were split into additional themes. For example, themes that only appeared in one set were often grouped under a broader theme.} We included a brief explanation for what each code represented. \revised{We then reviewed and revised this codebook to ensure that they consistently captured a useful range of ideas and experiences expressed by our participants}.


\begin{figure*}[htp]
    \centering
    \includegraphics[width=0.9\textwidth]{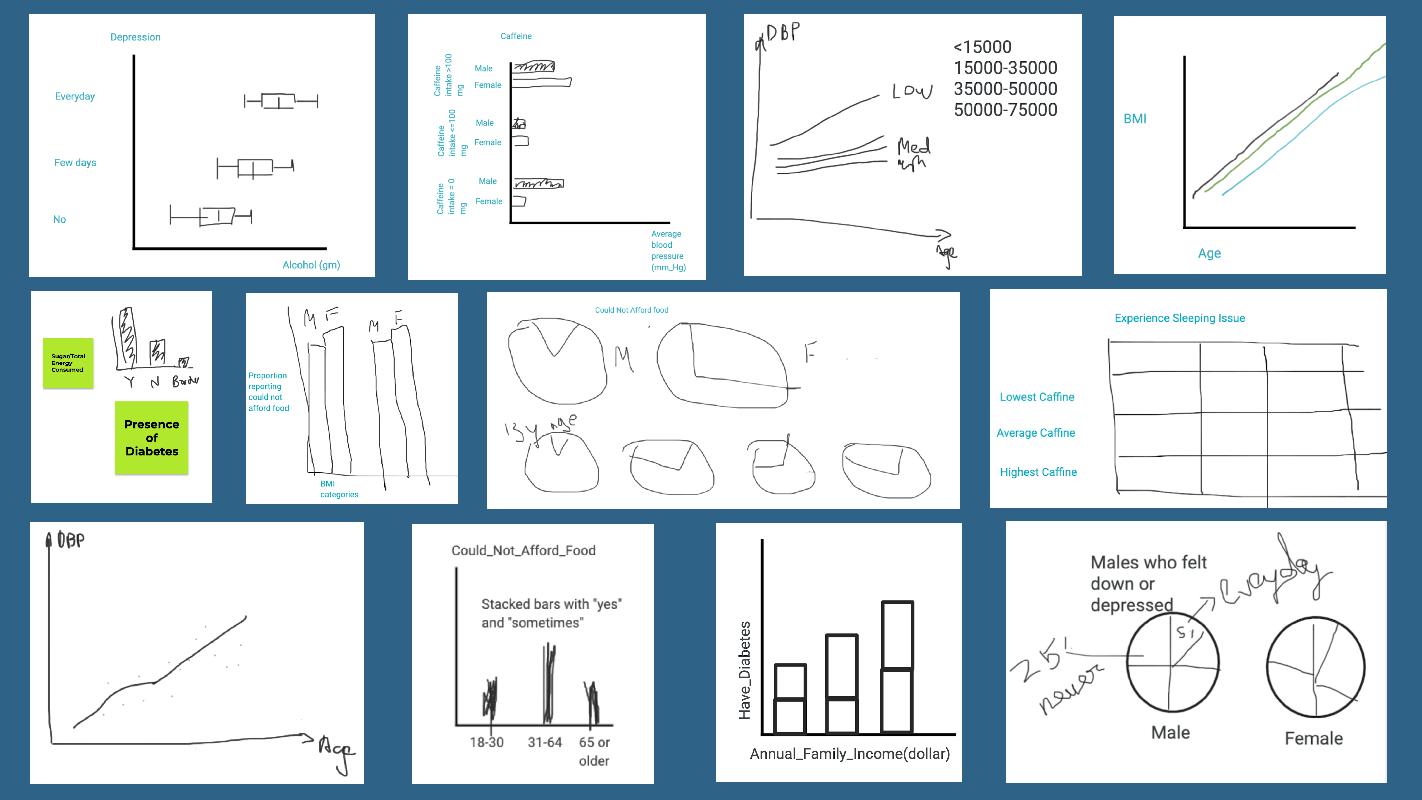}
    \caption{A gallery of recommendations sketched by participants.}
    \label{fig:sketch}
\end{figure*}

\subsection{Quantitative Analysis Methods}

Our general approach to quantitatively analyzing our study data was to programmatically extract the attribute selection, data transformation, and encoding selection choices made by participants using the Vega-Lite specifications derived from each visualization observed in our interviews. Given these extracted parameters, we counted observations of each choice, such as the total participants that created visualization recommendations that use bars as the mark type or the total participants that include color encodings in their created visualizations. Here, we describe how various parameters were extracted from our study data for analysis.

\subsubsection{Analyzing Recommendations Created By Participants} We translated each visualization recommendation created by participants into a corresponding visualization specification in Vega-Lite~\cite{satyanarayan2017vega-lite}. We created a series of Python scripts to analyze these specifications to extract different features, such as which attributes were selected from the \emph{NHANES} dataset, what data transformations were applied (e.g., binning and aggregation), and what mark types and encoding channels were chosen to visualize these attributes. We supplemented this automatic information with manual assessments of certain properties of the designs (e.g., which variable in a bivariate chart was the independent variable and which was the dependent variable).

\subsubsection{Analyzing Participants' Rankings of Existing Recommendations} Participants' rankings of visualizations from the pre-defined recommendation gallery were stored as a single relational table that recorded which 5 visualizations were selected from the gallery and each participant's ordering of their top five recommendations. Each visualization from the gallery was also translated into a corresponding Vega-Lite specification, with a note capturing recommendation source (human-generated, or auto-generated by Voyager or Dziban). We used the Vega-Lite specifications and the rankings table as inputs to scripts that extract the attributes, statistical patterns, visualization types, and encodings participants preferred from the gallery.

\subsection{What do Participants Prioritize When Creating Their Own Visualizations?}

In this section, we address question \textbf{RQ1} from \autoref{sec:motivation}: \emph{What characteristics of a visualization design do analysts prioritize when recommending them to colleagues?}

\begin{figure}[htp]
    \centering
    \includegraphics[width=0.95\columnwidth]{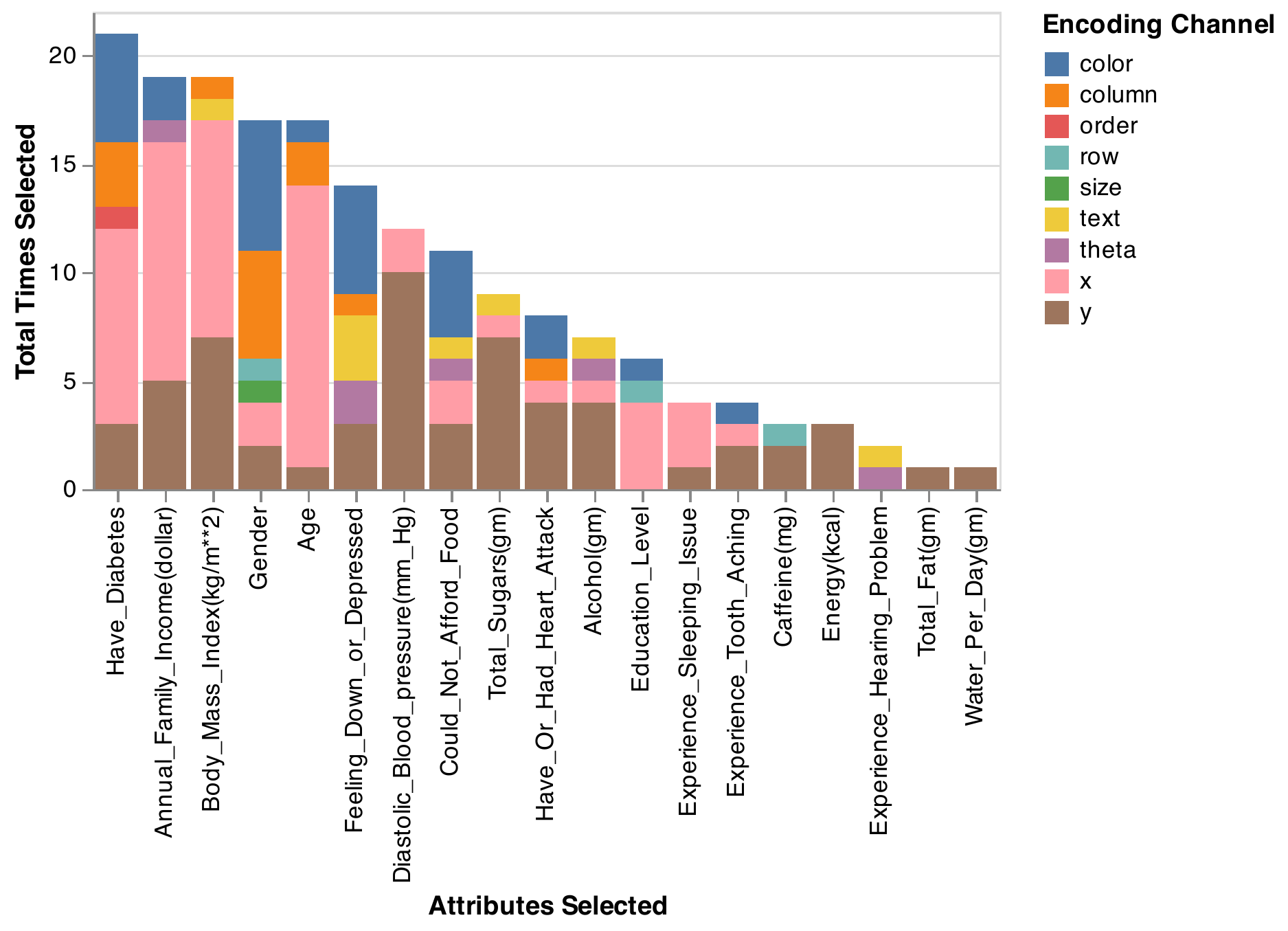}
    \caption{The frequency with which individual attributes from \emph{NHANES} were used in participants' sketched visualization recommendations from the Visualization Ideation Phase, colored by how each attribute was \revised{visually }encoded within each sketch. Positional encodings (such as x, y, row or column) were the modal ways that fields were mapped to visual channels.}
    \label{fig:encodings-attributes}
\end{figure}



\subsubsection{Participants created simple visualizations with a small set of intended takeaways or messages.}
\label{sec:analysis:simple-vis}

\begin{figure}[htp]
    \centering
    \includegraphics[width=5cm]{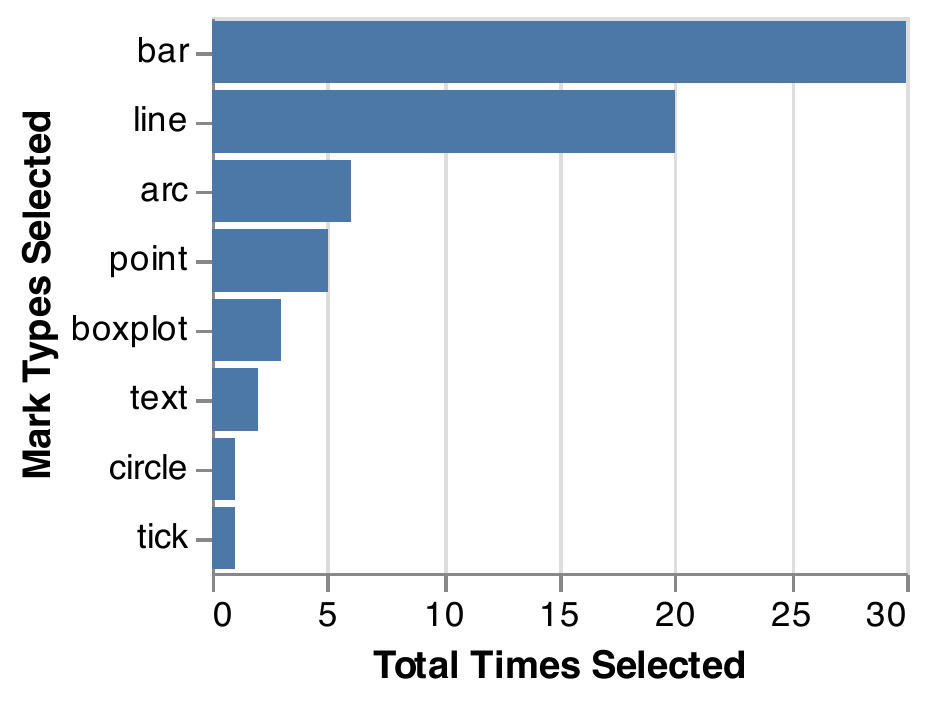}
    \caption{Visualization Ideation: Mark Representations}
    \label{fig:specs-marks}
\end{figure} 

In terms of the total attributes involved in each sketch, the vast majority of sketches include only two attributes from \emph{NHANES} (37 out of 53). A small subset includes three attributes (15 out of 53), and only one example incorporated five attributes. Most recommendations appear to favor bivariate relationships between data attributes, though multiple calculations were sometimes applied to individual attributes, producing more than one encoding for this attribute within a single visualization recommendation.

\revised{We did not name Tableau as the system used in our interview protocl, nor did we train or inform our participants on the specific capabilities of Tableau. Our participants were} not given any restrictions on the visualizations they could request. Nevertheless, participants eschewed ``xenographics''~\cite{xenographics} and stuck to common visual forms like bar charts, line charts, pie charts, and scatterplots (see \autoref{fig:specs-marks}). We did not encounter any visualizations that could not be converted to the intentionally limited Vega-Lite~\cite{satyanarayan2017vega-lite} specification language. While participants were often aware of more complex ways of showing the data, ease of interpretation was a frequent motivator for keeping things simple. For example, \textbf{P13} said ``\textit{a lot of my clients aren't really data-savvy. So I use a lot of bar charts, because they are pretty easy for them to understand... if you get too crazy or too fancy, then I lose them.}''

To further simplify their visualizations, participants generally applied aggregation to a dependent variable in their sketches to highlight distributions, trends, or patterns in the data. Examples include a boxplot visualization sketched by \textbf{P1} to show the distribution of alcohol consumption on the x-axis (dependent variable) grouped by whether respondents felt down or depressed on the y-axis (independent variable), or a bar chart sketched by \textbf{P15} showing the percentage of respondents with diabetes on the y-axis  (dependent variable) grouped by defined income classes on the x-axis (independent variable). Participants would further streamline the narrative of their visualizations by hiding or removing ambiguous records. For example, at least 9 participants wanted to de-clutter their visualizations by filtering out values such as ``Don't Know'', ``Refused to answer'', and nulls. \textbf{P12} tells the visualization mediator, ``\emph{Let's filter those out and just put it in a note. Yeah, I think three [categories] in a dataset this large might be distracting to include in the graph.}''

Participants also wanted clear take-aways or other action items for the intended audience, and would express disappointment when the visualization they sketched did not produce the expected clear trend. For example, when discussing all of their generated plots, \textbf{P1} was concerned that ``\textit{I don't think they're very actionable right now.}'' Similarly, when discussing a potential chart of BMI,  \textbf{P2} said ``\textit{I probably wouldn't send this to anybody because it looks the same across education levels, so that clearly doesn't matter much.}''

Finally, participants would use annotations or other design tweaks to highlight the intended message. \textbf{P11}, looking over a messy scatterplot, said ``\textit{I ... added a trendline to indicate some kind of trend so it's easier for the audience, especially a policymaker, to see the key message.}'' When deciding whether or not to aggregate their data, \textbf{P8} remarked ``\textit{If there's a clear trend, then that, to me, is more meaningful than on a scatter plot. That [scatter plot] doesn't really show anything.}'' \textbf{P8} also used titles to indicate intended messages (e.g. ``No trend between Diabetes status and Daily Alcohol intake'' and ``Higher Frequency of Toothaches is Associated with Higher Daily Sugar Intakes''), a design choice also undertaken by \textbf{P13}, who remarked ``\textit{I also tend to put in some kind of interpretive title or subtitle, so I tell them what they're seeing. So they don't have to guess.}''


\subsubsection{Participants focused on the visual appeal and legibility of their charts.}

Participants would often perform several rounds of iteration with the \revised{wizard} in the ideation task in order to improve the aesthetics and legibility of the final charts. \textbf{P8} mentions of their bar chart: ``\emph{I like putting outlines around the bars, so they stand out a little bit more... I also usually take off the lines going across}\footnote{Referring to the default background gridlines produced by Tableau Desktop}, \emph{because if you have the numbers on top of the bars, then you don't need the lines going across like that.}''). Some participants, like \textbf{P18}, conflated alternative mark types with aesthetic adjustments, mentioning: ``\emph{I would keep picking at these... for a long time to keep refining them and refining them... And sometimes, I'll make a figure as a bar chart, and then I'll just do a stacked bar... And then I'll try it as a pie chart. And I'll look at them side by side and decide,}'' an indication that iteration on visualization designs (including tweaks such as encoding channels, mark types and even miscellaneous adjustment) serves an important part in gaining fuller confidence in their presented visualization recommendations. The lack of ability to iterate on or control the design aspects of visualization designs may have contributed towards participant's reluctance to rank exterior recommendations more highly than their own creations (see \autoref{sec:analysis-rankings}). 

\subsubsection{Participants focused on data attributes connected with existing contexts and shared expectations.}
\label{sec:analysis:simple-story}

\autoref{fig:encodings-attributes} lists all of the attributes that participants used in their visualization sketches from the ideation section of our study (see \autoref{sec:experiment-design:experiment-flow}). 19 out of 20 attributes were used. The only attribute that participants did not use in their sketches was \texttt{Used\_Marijuana\_Or\_Hashish}. For the attributes used in multiple sketches, we found that \texttt{Age} and \texttt{Education\_Level} were the only attributes used exclusively as independent variables across all sketches. Similarly, \texttt{Total\_Sugars} was the only attribute to be used exclusively as a dependent variable. Otherwise, attributes appeared both as independent and dependent variables within visualizations. For example, a participant might perform a breakdown of two demographic variables in either order (say, \texttt{Have\_Diabetes} by \texttt{Age}, or \texttt{Age} by \texttt{Have\_Diabetes}), placing the variable in a dependent or independent role depending on their specific analytic context (e.g. ``Are older people more likely to have diabetes'' or ``Are people with diabetes more likely to be elderly'').

That being said, most demographic measures were consistently used as \emph{independent} variables in participants' sketches, such as \texttt{Annual\_Family\_Income} (used as the independent variable in 12 of 16 sketches including the attribute), \texttt{Gender} (13 of 14), \texttt{Age} (11 of 11), and \texttt{Education\_Level} (4 of 4). Most health measures were generally treated as \emph{dependent} variables, such as \texttt{Body\_Mass\_Index} (8 of 14 sketches including the attribute), \texttt{Have\_Diabetes} (7 of 11), \texttt{Total\_Sugars} (8 of 8), or \texttt{Diastolic\_Blood\_Pressure} (5 of 7). Most sketches involved one demographic measure and one health measure, or two health measures. The assignment of causal roles for these variables was not arbitrary. From the rationales provided by participants over the course of the ideation exercise, participants mentioned the need to identify ``drivers'' and ``trends'' (\textbf{P1}) connected to health or health outcomes. As per \textbf{P3}, ``\emph{I thought about more social determinants of health, and what would be affecting it}'' and  \textbf{P13} remarks: ``\emph{I'm looking for things that relate more to getting to outcomes, whereas some of these seem more like physiological relationships or demographic relationships.}''

\begin{figure}[htp]
    \centering
    \includegraphics[width=0.4\textwidth]{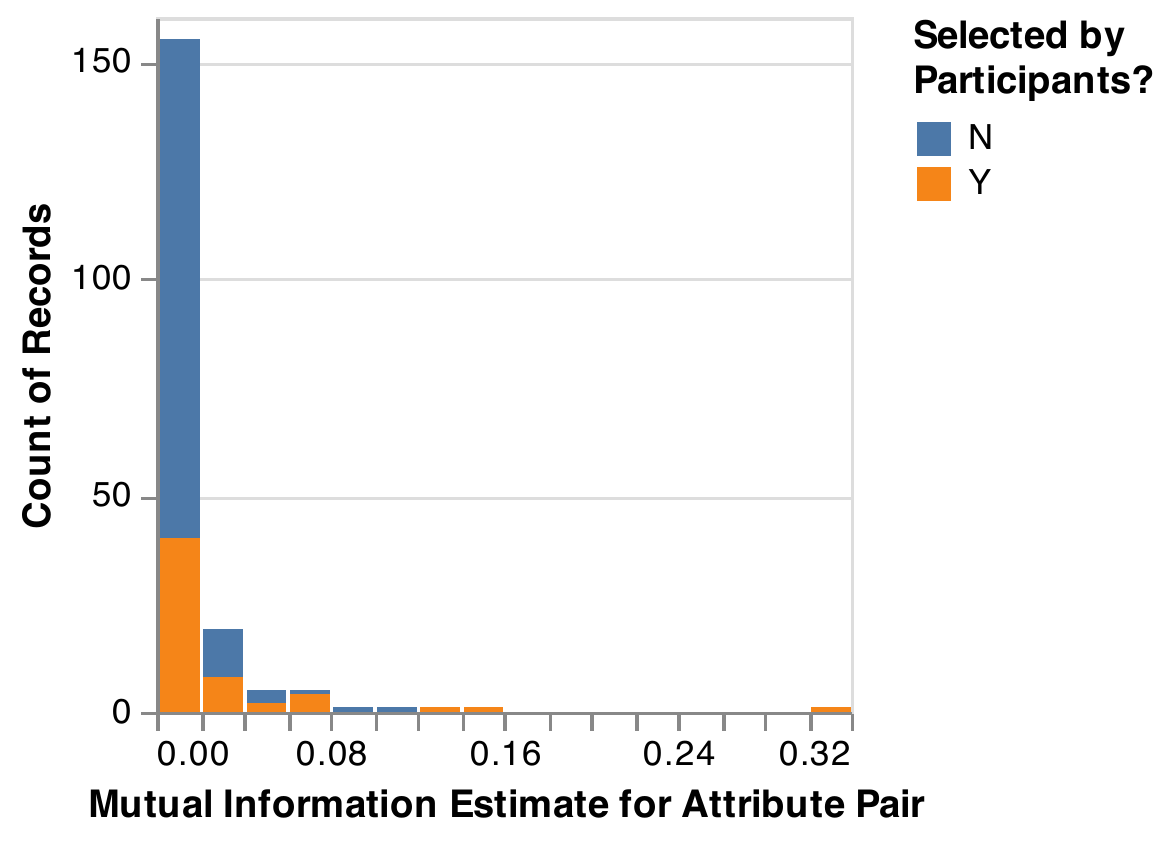}
    \caption{
    Counts for all possible pairs of attributes, binned by estimated mutual information value (similar to  correlation)~\cite{cover2012elements}. A higher mutual information estimate indicates a stronger correlation for the given pair of attributes. The pairings that were used in participants' sketches are colored orange.
    }
    \label{fig:mutual-information}
\end{figure}

The pairing of independent and dependent variables were typically made in advance of seeing the rendered results, suggesting that participants utilized prior knowledge and/or experiences to form hypotheses regarding potential correlations and causal relationships. That is, rather than looking for highly correlated fields and surfacing them to users, participants selected fields where the relationship was perceived as relevant or important, regardless of the resulting statistical correlation or (lack of) visible trend. For example, \textbf{P8}, upon seeing the results of a bar chart showing alcohol consumption broken down by diabetes status, remarked ``\emph{It doesn't show me anything, should I change it? But I mean, that's sometimes good to know to see that there isn't really a trend between two things.}''

To confirm the disconnect between statistical correlations and the trends our participants chose to surface, we estimated a mutual information value for all possible pairs of attributes in our \emph{NHANES} dataset (similar to a correlation score)~\cite{cover2012elements}. Then, we identified which pairings were actually observed in participants' sketches, which are colored orange in \autoref{fig:mutual-information}. A higher correlation between the values in the selected attributes translates to a higher mutual information estimate. We observed a slight preference for higher ranking pairs, however, pairings span the full spectrum of observed mutual information scores.

Providing recommendations that were linked thematically or otherwise mutually supportive was also an important factor for participants, who preferred to keep a common theme among their generated visualizations. For example, \textbf{P11} (interest in \texttt{Diastolic\_Blood\-\_Pressure} on the y-axis for all three plots), \textbf{P12} (interest in \texttt{Have\-\_or\-\_Had\-\_Heart\-\_Attack} in all three plots), \textbf{P13} (interest in \texttt{Have\-\_Diabetes} for all three plots), and \textbf{P3} (interest in \texttt{Total\-\_Sugars} and \texttt{Total\_Energy} in two plots) all preferred similar data attributes and/or plot types when sketching their three recommendations in order to create a synergistic, mutually supportive dashboard.  



\subsection{What do Participants Prioritize When Ranking Pre-Defined Visualizations?}

In this section, we address question \textbf{RQ2}: \emph{What do analysts prioritize when evaluating visualization recommendations from other sources?}

\subsubsection{Participants preferred simplicity over information overload.}
When presented with ten visualizations, participants vastly preferred simple graphs that could be interpreted easily. While ranking the visualizations, \textbf{P1} mentions ``\emph{I think some of the graphs had a little more information than what might have been easily digestible.}'' and similarly \textbf{P3} ranked the visualizations ``\emph{mostly by simplicity: which ones were quickest and easiest to follow?}.'' Participants also commented on the ambiguous data present in various visualizations in the gallery, also discussed in \autoref{sec:analysis:simple-story}. \textbf{P13} mentions \say{\emph{Like some of the things just confuse your data a little bit. Like we still have nulls here. We have `Don't Knows', we have `Refused', so like I'd probably take those out if I was doing this.}}

As with the ideation task (\autoref{sec:analysis:simple-story}), desire for simplicity \revised{was} also reflected in a preference for visualizations with clear takeaways. \textbf{P4} reports ``\emph{trying to prioritize one set... where the trends jumped out a little bit more and were easier to look at to get what it's trying to tell you at face value. That was the number one thing.}'' When presented with a visualization without a ``clear trend,'' \textbf{P9} argues ``\emph{If it does not have a lot of variation (in the graph), then why do we do it this way?}'' These sentiments echo participants' preferences from the ideation task, where many participants would scrap an idea if it did not lead to a visualization with positive results.

Our participants valued simplicity (and avoided complexity) in their recommendations for a variety of reasons:
\begin{enumerate}
  \item The participant was not able to understand a few of the relatively complex visualizations and would not recommend them to others. For example, \textbf{P5} had difficulty interpreting one graph and said ``\emph{like this, I find really hard to understand. To me, you have to look at too many things and then make sense of what it's telling you.}''
  \item The participant determined that the complexity of a visualization \revised{would} hinder \revised{its} effectiveness for communicating the story to the target audience (policymakers). \textbf{P15} explains ``\emph{And, some of these I think are just so complicated for people that understand what they are so I'm not choosing them. I'm also just trying to think about what would be easy for my client to show and what people would get.}'' Referring to the same visualization (Visualization ID A), \textbf{P6} liked it because ``\emph{...it may be a little less fancy or something like that, but it communicates the information pretty easily, which is always good.}''
\end{enumerate}

Our quantitative analyses reinforced our qualitative findings here. In terms of encoding count, we found that participants preferred recommendations representing bi-variate relationships over more complex visualizations: 48.9\% of participants' selected gallery visualizations contained exactly two attributes, 41.1\% of participants' selected gallery visualizations contained three attributes, sharply dropping off to 6.7\% for gallery visualizations containing four attributes. 

\subsubsection{Recommendations were often considered inspiration for later exploration but not necessarily a final product.}\label{sec:explr-not}

Through ranking the pre-defined gallery of recommendations, participants discussed the things they liked and disliked for each visualization. Several considered the visualizations interesting which helped spark ideas that they had not come up with while generating their own visualizations. \textbf{P5} was inspired by one area chart, and wanted to explore it further. ``\emph{I think it could be super cool. But as an area chart, I guess I'd have to play around with the data to see.}'' \textbf{P1} noted that they'd like to dive into the recommended visualizations to look at subsets of the graph: ``\emph{I think I would like take cuts, like look at a subset of people who are feeling down and depressed, and then really analyze that.}''  



These findings suggest that the visualization recommendation process should be treated as a dynamic, evolving dialogue between the user and the system, where recommendations can spur new ideas but not necessarily substitute for the user's own visualization design work.


\subsection{How Do Participant Recommendations Compare With Those From Other Sources?}
\label{sec:analysis-rankings}

In this section, we address question \textbf{RQ3}: \emph{How do the recommendations made by analysts align with those created from other sources in terms of visual form or analytical purpose?}

\subsubsection{Participants preferred their own creations to recommendations from others.}


Only 6 out of 18 participants picked gallery visualizations to replace their own created visualizations. Our observations suggest several potential reasons that together point to a participants' perceived difficulty parsing the visualization, including: aesthetics \& visual formatting (``\emph{This one is sort of hard to read since the x-axis isn't labeled and that just sort of hard figuring out what it's saying.}'' -\textbf{P12}), and simplicity or clarity (``\emph{I rarely make a visual for a technical audience, and so I look at these charts and I'm like, my client wouldn't understand this.}'' -\textbf{P5}). These factors were occasionally valued over even analytic utility or accuracy, e.g., ``\emph{Some of these graphs, like A or D, might actually be describing the dataset better than the ones I selected. But because it's so much harder to read... that's not something I would show a client. That's why I didn't even go further than an initial look.}'' (\textbf{P19}). Together, these findings suggest that analysts are consistent in their recommendation preferences---in this case, prioritizing clear and simple narratives---regardless of whether they are creating new recommendations or ranking existing ones.

\subsubsection{Participants exhibited no strong preferences for recommendations generated by other humans versus those generated by recommender systems.}

In terms of recommendation sources (e.g., human versus algorithm), we found that when selecting from the gallery, participants had a slight preference for human-curated recommendations (48 selections) over those generated using Voyager and Dziban (24 and 18 selections, respectively -- 42 total).
However, we note that our piloting process had, in a way, already pre-filtered algorithmically generated recommendations (see \autoref{sec:experiment-design:pilot-study}) based on interest. We are therefore hesitant to use our quantitative results to point to a strong pattern of preference for a particular source of recommendation, beyond pointing out that there was no extreme bias towards one source of recommendations or another.

\revised{
\subsection{Reflections from the Interviewers}

Our original preregistration focused on determining how our participants perceived the recommendation design process, but did not consider the perspective of the interviewers (the mediator and the wizard). In this section, the interviewers in the experiment reflect on their experiences and observations from the interview transcripts. We deviate from our original preregistration to include these perspectives as we believe they provide richer nuance surrounding opportunities and challenges within the visualization recommendation process.

We highlight notable misalignments in expectations that we observed throughout our study, clustered according to three recurring themes: uncertainties about the data from the user's perspective, uncertainties about user intent from the wizard's perspective, and the inability to generate the visualization according to the original user intent. These misalignments led to interesting compromises in the end-visualization, as detailed below.

\subsubsection{Assumptions and Uncertainties about Data}
On several occasions, the wizard interfered with the participant's design process by reminding them of characteristics of the data. However, on one occasion, this interruption spun into the participant selecting a radically different choice in data attributes.

\begin{quote}
    \begin{itemize}[label={}]
        \item \textbf{Wizard}: For the second line plot where you had income by BMI... One thing to keep in mind is that income isn't actually a continuous variable; it's discrete with a size of 5k.
        \item \textbf{P6}: Thanks for pointing that out, I did not notice that. Let's do something else then... Let's just ask a curious question. Instead of the BMI and salary, let's do caffeine, by age and by gender.
    \end{itemize}
\end{quote}

While in some instances (such as the above) this new awareness or clarification about the data caused the participant to change plans entirely, in the vast majority of scenarios, the wizard and the participant worked to derive meaningful compromise on the data attributes once the data was clarified. For example, this snippet from the \textbf{P18} transcript shows an incremental back-and-forth which fleshed out a visualization design, as gaps and uncertainties in knowledge of the dataset were resolved between the wizard and the participant:
\begin{quote}
    \begin{itemize}[label={}]
        \item \textbf{Wizard}: Just a quick note that for the first graph you sketched, ``Could Not Afford Food'' actually can contain three different values.
        \item \textbf{P18}: Thank you for pointing that out... I shouldn't have assumed it was [boolean].. Can you do it as like a stacked bar [instead of a standard bar chart]?
    \end{itemize}
\end{quote}

In addition to highlighting the value of mutual communication and iteration when creating visualizations, these interactions also suggest the utility in presenting users with useful summaries or alerts about the data available to them in order to orient them to their data.

Given the shape and size of our chosen dataset, generating many standard visualizations required an explicit choice of aggregation. Our own participants often did not explicitly qualify an aggregation method or took it for granted that the system would use their intended default. This likely extends to when they interact with drag-and-drop interfaces such as Tableau. In the experiment, the wizard had to thus frequently prompt for explicit aggregation instructions, such as in the \textbf{P16} transcript: 

\begin{quote}
    \begin{itemize}[label={}]
        \item \textbf{Wizard} [referring to bars, after \textbf{P16} finished sketching]: So would it be like... the average sugar intake?
        \item \textbf{P16}: Mmm.. Yeah! You could do that [averaging] for any of the health outcomes and see your relationship with annual family income.
        \item ... 
        \item \textbf{Wizard} [referring to a second sketch]: For the y-axis, did you want to zero in on one attribute in BMI-sugar-caffeine? Should we just pick one?
        \item \textbf{P16}. Sure. BMI, although that seems like probably well-established already in a policy.
        \item \textbf{Wizard}: Okay, so for now, I'll do BMI. Do you have an aggregation measure you'd like to pick? What would you like to do? I'd like to remind you that in the first graph, we decided to average the total sugar intake.
        \item \textbf{P16}: Again, to average the amount.
    \end{itemize}\label{Clarification on intent for visualization marks}
\end{quote}

This uncertainty around aggregation type has been observed not just in creators but also readers of visualizations~\cite{kim2019designing} and suggests that choosing the wrong ``default'' could produce confusing or even misleading recommendations.

Lastly, the participants often asked questions that required deep knowledge of either the domain or the source data. As the wizard and the mediator were not public health experts themselves, these questions occasionally led to acknowledgments of uncertainty:

\begin{quote}
    \begin{itemize}[label={}]
        \item \textbf{P11:} For the diabetes question, how was that question worded?
        \item \textbf{Mediator:} Regarding the exact question, I can't recall off the top of my mind, but the user can answer either "Yes", "No", and "Borderline".
        \item \textbf{P11:} Okay, so the question didn't mention anything about diagnosis?
        \item \textbf{Mediator:} I'm not 100\% sure.
    \end{itemize}
\end{quote}

Some of these details about the dataset matter and affect how a participant would generate a visualization. We did not always have the answers to some of these questions. In a similar way, many visualization systems do not have all of the semantic details about a dataset it might visualize for a user, though this would streamline the design process by improving the user's literacy on the dataset and improving their ability to find appropriate visualization forms for their data.

\subsubsection{Uncertainties About User Intent}\label{sec:analysis-user-intent}
Analogously to the participant's uncertainties about the data, there were sometimes uncertainties on the wizard's perspective about user intent. This sometimes led to scenarios that even after a couple clarifying questions, we were unable to 100\% follow what a user expected. This was often compounded when the participant would provide vague visuals.

\begin{figure*}
\centering
\begin{subfigure}{.8\columnwidth}
  \includegraphics[width=\linewidth]{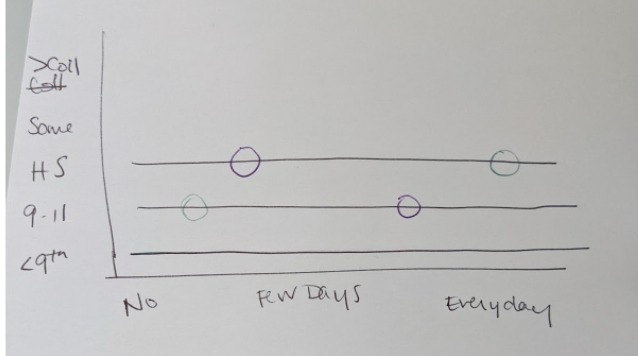}
  \caption{\textbf{P5} sketch. \textbf{P5} notes that nodes are colored according to the dataset's gender attribute.}
  \label{fig:p5_sketch}
\end{subfigure}
\hfill
\begin{subfigure}{.8\columnwidth}
  \includegraphics[width=\linewidth]{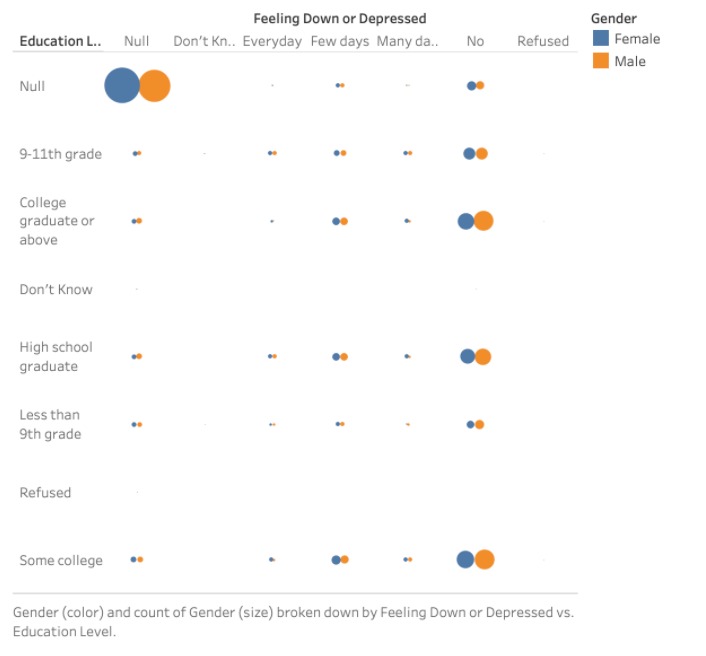}
  \caption{Wizard's interpretation of \textbf{P5} sketch}
  \label{fig:wizard_gen_p5}
\end{subfigure}
\caption{\textbf{P5}'s initial design idea (\autoref{fig:p5_sketch}) and the wizard's interpretation of the design (\autoref{fig:wizard_gen_p5}). These visualizations have slight differences, including filtered values on both axes, as well as more structural differences such the data grid inherent in \autoref{fig:wizard_gen_p5}. For this sketch, the wizard was unclear on the design intent of \textbf{P5}, specifically regarding the mark types, which ultimately led to their inability to specify this in Tableau.}
\label{fig:side-by-side-p5}
\end{figure*}

\begin{quote}
    \begin{itemize}[label={}]
        \item \textbf{P5:} So I don't know how doable this is, I would love to, and there may not be time for this for like what you have planned for today... So I'm thinking like, if we had more time, it would be cool to group education level differently. Right. So like, anyway, so like, less than ninth grade, and ninth to 11th grade could be grouped as one, but obviously, we don't need to do all that grouping now. But like that grouping would just have two dots on its line. Right, it would show the x axis which would be the spectrum of no to every day, like no over at the zero, what up to essentially, the education grouping would have two dots on its line, as opposed to right now it has 10 you see what I'm saying? Yeah, it's far more simple. Or we can move on to the other one.
        
        \item \textbf{Wizard:} Okay, let's keep this for now. And then if we have time, we can go back and do these adjustments.  
    \end{itemize}
\end{quote}

There was also at least one case in which the wizard was unable to meet the participant's needs with an alternative visualization design.

\begin{quote}
    \begin{itemize}[label={}]
    
        \item \textbf{Wizard:} Oh, yeah, that um, that is up to you if you have a different way that you want to visualize it. Since the table is a little bit tricky. 
        
        \item \textbf{P20:} So I think then we can go for line graph also. 

        \item \textbf{Wizard:} Okay. Yeah, so I think for for this one, we have a line graph, it can't be done since the there are categories for the body, since it's because of the different categories. So let me show you. Like we can have this bar chart with the annual family income versus each of the different types of different types of BMI. Yeah, so is this an alternate solution? 

        \item \textbf{P20:} Okay. This is still not quite what I am looking for.
        
        \item \textbf{Interviewer:} Right, I guess, if that's the case, we can still mark that you still want the table? Because I realized we're like overtime a little bit... Yeah, we can definitely take notes on that and maybe mark that down in our notes. 
    \end{itemize}
\end{quote}

These moments ultimately affected the wizard's ability to address or tweak user intent in the design. While having a \emph{wizard} in the loop was in some sense detrimental for full expressiveness (since the wizard may have had imperfect knowledge of the system), it also afforded us fuller \textit{transparency} in communicating both system and wizard limitations to users, as well as \textit{mindfulness} on the part of the participants of potential designs that required significant time or effort to ``get right.'' That being said, it is presently difficult to provide the same expressiveness a user might have with pen-and-paper to a visualization system, especially when factors such as data, domain, and technical know-how exist. Compromises must occur, which is why we detail these moments in our interviews here.

\subsubsection{Constraints: System and Wizard Ability}

While the wizard tried to maintain the original user intent as intact as possible, this wasn't always possible due to constraints on the system (Tableau) or the wizard's ability to use the system to create a visualization isomorphic to the participant's initial design, in the time allotted. Therefore, while we always recorded the original sketch and design intention of the participant, on several occasions, the wizard and the user were forced to produce a compromised design, or abandon potential design ideas, sometimes due to the request for more advanced analytical queries.

\begin{quote}
    \begin{itemize}[label={}]
        \item \textbf{Wizard:} So let me try to understand what you were looking for. So you're looking for a proportion graph per income level, where the number essentially represented is the numerator [family income] divided by the total number of respondents across all family incomes?
        \item \textbf{P9:} No. So like, let's say there are 100 respondents in the 5k category. 100 males and 100 females. I want to see of those 100 males, what proportion reported feeling down or depressed? Maybe it's only 10\%? And then among females, what proportion reported feeling down or depressed?
    \end{itemize}
\end{quote}

Though we tried to limit drastic interference (e.g. ``here's a radically different visualization design idea you could consider''), sometimes compromises due to technical skill with the visualization system had to be made.

\begin{quote}
    \begin{itemize}[label={}]
        \item \textbf{P9:} I'm just trying to imagine what that could look like, besides doing like side-by-side bars for the males and the females. Do you have any suggestions?
        \item \textbf{Wizard:} Well, we could just separate the axes and make one axis female and one male. 
        \item \textbf{P9:} Okay, yeah I like that. So female versus male.
    \end{itemize}
\end{quote}

\begin{figure*}[htp]
    \centering
        \begin{subfigure}{0.95\columnwidth}
          \centering
          \includegraphics[width=\linewidth]{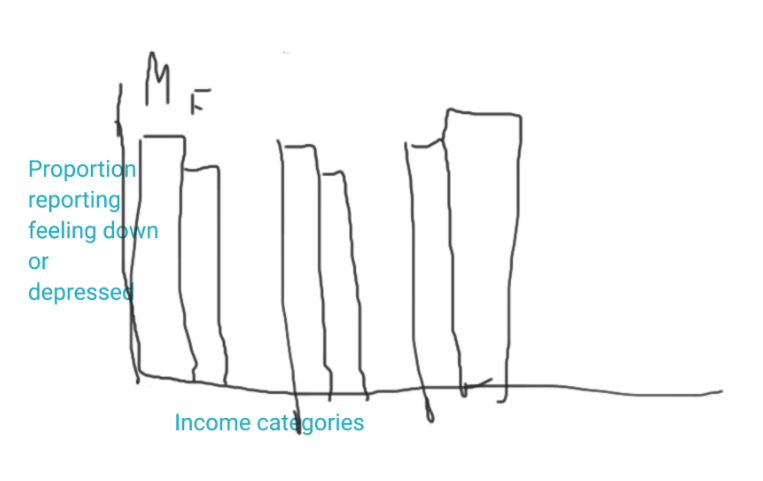}
          \caption{\textbf{P9} sketch.}
          \label{fig:p9_sketch}
        \end{subfigure}
    \quad
        \begin{subfigure}{0.95\columnwidth}
          \centering
          \includegraphics[width=\linewidth]{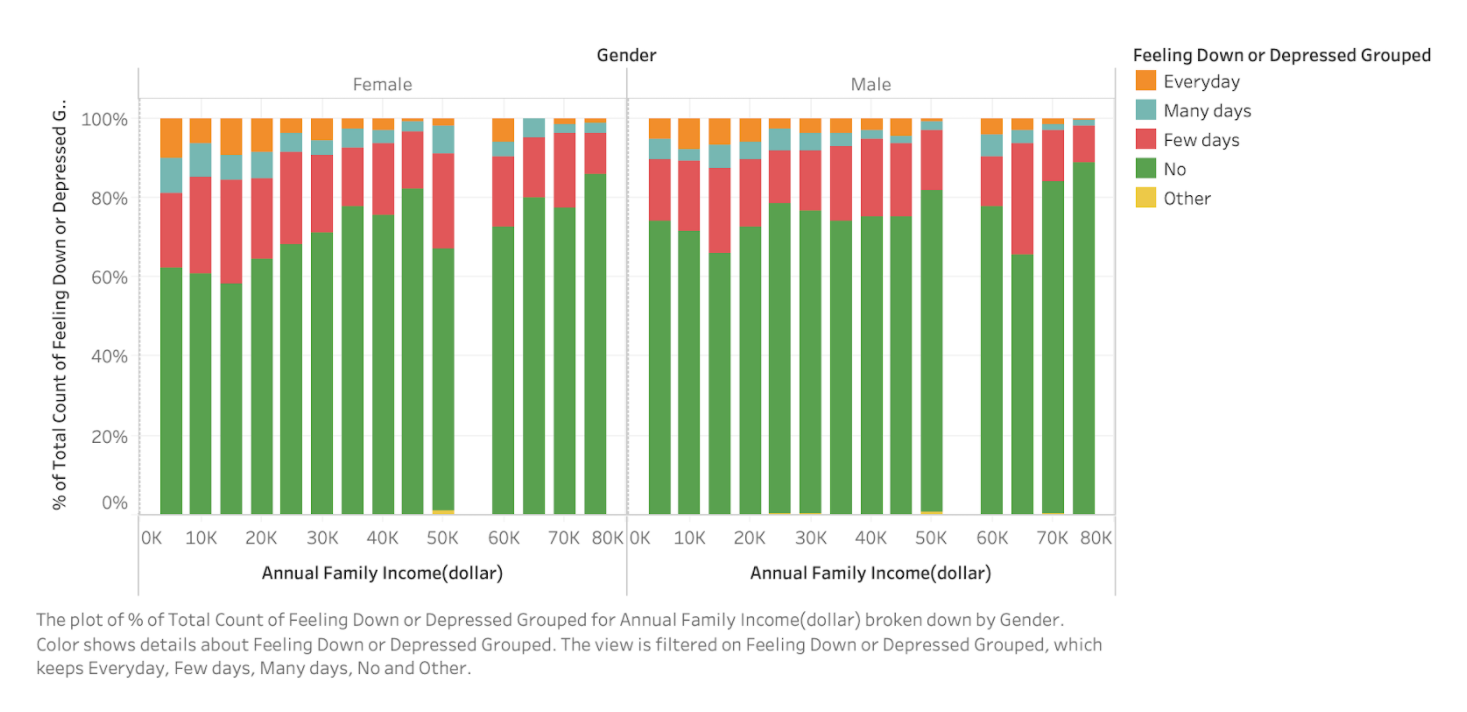}
          \caption{The wizard's interpretation of \textbf{P9} sketch}
          \label{fig:wizard_gen_p9}
        \end{subfigure}
    \caption{Participant \textbf{P9}'s initial design idea (\autoref{fig:p9_sketch}) and the wizard's interpretation of the design (\autoref{fig:wizard_gen_p9}). Despite the apparent simplicity of \autoref{fig:p9_sketch}, \textbf{P9} paired it with more advanced analytic functions that were too technically challenging for the wizard to implement in the time allotted. The design compromise is shown in \autoref{fig:wizard_gen_p9}.}
    \label{fig:test}
\end{figure*}


The examples in this section highlight that the wizard was not a real-time translator of a participant's visualization idea into a final design, but had to navigate their own technical skill with Tableau, along with the uncertainties and ambiguities regarding user intent and data. For all sketches, the wizard was able to generate a design compromise that the participant accepted for the sake of ``moving on.'' However, the scenario demonstrated in the snippet with \textbf{P9} may not translate to a real-world scenario where certain analytic functions are necessarily to be expressed in the visualization. Sometimes, the request of advanced technical functions on data attributes led to more creative, bespoke visualizations in which the wizard, though decently familiar with Tableau functions, was not technically prepared to reconstruct during the interview, hampering the participant's visualizations. Although we were mostly able to come to a compromise with the participant's visualization design, it becomes more difficult to manage their expectations for visualizations involving complex transforms, which may not be easily replaceable with a simpler one, when they are requested.
}

\section{Discussion}
\label{sec:discussion}

While there were many themes and potential vignettes related to our data, we focus our discussion on three recurring themes encountered in our analysis. These themes represent \revised{\textit{design values}} or \textit{priorities} that recurred across participants and across our experimental tasks: \textbf{simplicity}, \textbf{relevance}, and \textbf{interest}.
We use this section to explicate these \revised{design values}, with a focus on how these values might inform current or future designers of visualization recommendation systems. \revised{We reiterate that our choice of participants and methods results in knowledge that is positional, and exists in the context of our pool of public health researchers and analysts. What counts as ``simple, relevant, and interesting'' for them may not necessarily hold true for other domains or target audiences.}

\subsection{Visualizations Should Be Simple}

As discussed in \autoref{sec:analysis}, participants repeatedly preferred \textit{simple} visualizations with minimal attributes both when creating and ranking visualizations. This notion of simplicity extends to both the visual \textit{design} of the visualizations as well as the \textit{messages} of these visualizations. In terms of \textit{design}, this simplicity is associated with a preference for familiar visualization designs (bar charts, line charts, and scatterplots) encoding only one or two variables at a time. In terms of \textit{messages}, this simplicity was reflected in a preference for clear or otherwise unambiguous statistical patterns (such as expected trends or relationships between two variables), with filtering and aggregating used to remove extraneous information from a chart.

Our participants provided several rationales for preferring simplicity, but two of the most relevant for designers of recommendation systems were the desire to present clear takeways for an audience with variable experience interpreting complex or unfamiliar charts, and the sense that visualization recommendations should serve as a starting point for future analyses, rather than an ending point. From these \revised{design values}, we propose two guidelines for designers of recommendation systems. Namely, recommendation systems should:

\begin{enumerate}
    \item \textbf{produce charts from common genres with bounded visual complexity}. A visualization recommendation system, especially one for the purpose of orienting or familiarizing a user with a dataset, is not the place to try esoteric or unfamiliar charts. Similarly, rather than attempting to include every potential relevant dimension in one design, the recommendation system should value and present potentially greater sets of univariate or bivariate charts.
    \item \textbf{offer opportunities for refinement and extension}. Especially in the context of our suggestion above, a single ``simple'' visualization may not be sufficient to meet the analytical goals of all users. Rather than increase the complexity of the recommended chart to cover all potential goals, systems should allow users to modify or extend recommendations within more traditional exploratory data analysis workflows. Rather than an ``automatic insight,'' a recommendation might be better thought of as an interesting question or hypothesis to verify in more detail or with more nuance after the fact.
\end{enumerate}

\revised{Note that a person's perception of simplicity can also be influenced by their prior experiences with certain encoding channels, which is dictated in part by their discipline. For example, domain experts may prefer to use established encoding conventions~\cite{brehmer2016matches} from within their discipline, even though these encodings may technically be more complex or less legible~\cite{dasgupta2020effect}. Additional studies could prove beneficial in distinguishing domain-specific guidelines from design principles that could apply to any type of visualization recommendation system.}

\subsection{Visualizations Should Be Relevant}

As mentioned in \autoref{sec:related-work}, many visualization recommendation systems are built to be agnostic to the data domain, providing recommendations purely based on design guidelines, statistical features, or the syntactic structure of the data. Our participants, however, were more motivated by the \textit{anticipated needs} of their audience and their own \textit{structural and semantic} understanding of the data. As with simplicity above, we acknowledge that \textit{relevance} is similarly polysemic and nuanced as a concept.

For instance, participants often zeroed in on specific variables \textit{a priori}, excluding variables they felt to be irrelevant to their likely audience or unrelated to another pre-selected variable of interest. These choices represent implicit or explicit \textit{hypotheses} about the relations in the data, as well as implicit or explicit \textit{causal modeling} of how one factor might influence another. Providing evidence for or against these implicit hypotheses resulted in a common recipe for bivariate visualizations: a demographic factor cast as an independent variable, with a health outcome measure cast as a dependent variable. 

The centrality of domain relevance leads us to make the following recommendations for designers of visualization recommendation systems. Namely, systems should:

\begin{enumerate}
    \item \textbf{incorporate domain contexts and field relationships}. This could be as simple as giving common fields like \textit{Age} or \textit{Gender} fixed roles (say, as independent variables) in recommended charts, or as complex as attempting to infer or solicit causal graphs, functional dependencies, or other field relationships in particular datasets. This recommendation also suggests that the promise of a \textit{universal} ``auto-insight'' system is potentially misguided, or at least a lofty dream, and systems seeking to present relevant data may need to be tailored to specific domains or users.
    \item \textbf{allow users to specify intent or their analytical goals}. The exact same visualization may be alternatively useful or irrelevant across different users depending on their interests or prior assumptions. Allowing users to specify particular fields of interest or even their hypotheses and assumptions (see below) would allow systems to be more adaptable and retain relevancy across users or analytics sessions. While Q\&A systems (see \autoref{sec:related-work:qa}) have begun to incorporate models of intent into their design, we believe that this sort of contextual or ancillary semantic information is critical across all sorts of recommendation systems.
\end{enumerate}

\subsection{Visualizations Should Be Interesting}

The last \revised{design value} we examine in the context of our findings is that of preferring \textit{interesting} visualizations over visualizations that gave the viewer (or the audience) no new and useful information. This need for a visualization to show \textit{something} (a trend, a skew, or just the more general notion of a ``takeaway'') has potentially troubling implications for system designers.

A common pattern was for participants to abandon (in our ideation task) or downweight (in our selection task) visualizations without definitive patterns and trends. This includes visualizations showing negative or null results with respect to an assumed trend or difference between or among groups. \textbf{P8}'s verbal quandary (see \autoref{sec:analysis:simple-vis}) over whether to change a visualization that ``doesn't show me anything'' was an exception in this respect. In other words, critiques of auto-insight systems (see \autoref{sec:related-work:autoinsight}) based on their potential to highlight dramatic but spurious effects may be similarly applied to human recommenders. While we therefore \textit{could} suggest that recommendation systems present only the most salient of positive results, as an alternative we present the following suggestions. Namely, that systems should:

\begin{enumerate}
    \item \textbf{allow users to specify explicit hypotheses or assumptions}. A chart showing absolutely no correlations or clear patterns might still be of interest to a viewer with a strong expectation or assumption about a relationship in the data. To afford these sorts of interesting findings, recommendation systems (especially auto-insight systems) should embrace what Hullman \& Gelman~\cite{hullman2021designing} refer to as ``theories of graphical inference'' such that visualizations can provide direct evidence for or against inferences about the data. Kim et al.~\cite{kim2017explaining} in particular suggest  externalizing expectations and predictions prior to looking at the data as an alternative to traditional data presentation.
    \item \textbf{provide additional guidance to users on the robustness or importance of the visual pattern in a particular recommendation.} To prevent users from being misled by potential spurious patterns in sets of visualizations (such as those encountered and reported by the participants observed by Zgraggen et al.~\cite{zgraggen2018investigating}), recommendation systems may need to provide additional information on the reliability of findings (either through existing or novel metrics of reliability, or through alerts or warnings of potential threats to validity~\cite{mcnutt2020surfacing}).
\end{enumerate}



\section{Conclusion}
This work explored the preferences and priorities of in-domain analysts when creating and evaluating visualization recommendations for their target audience. We base our findings off semi-structured interviews with 18 analysts in the public health sector, observing behaviors, attitudes, and perceptions they had for various components throughout the visualizations in the experiment. Our findings highlight that these users overwhelmingly value \textbf{simplicity, relevancy, and analytic interest} both when creating their own visualization designs and when evaluating other visualization recommendations. Participants either demonstrated an understanding of a visualization and the potential story or insight stemming from it, or were able to formulate targeted questions to any uncertainties with the attributes in the visualization. Furthermore, certain data attributes (demographic and health outcomes) were frequently paired in visualization designs, indicating natural priors and biases in semantic knowledge in the dataset. Lastly, we find that participants more likely engage with visualizations showing seemingly ``positive'' results with perceptively clearer patterns or trends. Based on our findings, we suggest various design possibilities for visualization recommendation designers that could better aid in data exploration workflows.
\begin{acks}
We thank all the reviewers, study participants, and members of both the Human-Computer Interaction Lab and the Battle Data Lab for their valuable feedback.
\end{acks}

\bibliography{references.bib}

\end{document}